# HoneyDOC: An Efficient Honeypot Architecture Enabling All-Round Design

Wenjun Fan, Zhihui Du, Senior Member, IEEE, Max Smith-Creasey, and David Fernández

*Abstract*—Honeypots are designed to trap the attacker with the purpose of investigating its malicious behavior. Owing to the increasing variety and sophistication of cyber attacks, how to capture high-quality attack data has become a challenge in the context of honeypot area. All-round honeypots, which mean significant improvement in sensibility, countermeasure and stealth, are necessary to tackle the problem. In this paper, we propose a novel honeypot architecture termed HoneyDOC to support all-round honeypot design and implementation. Our HoneyDOC architecture clearly identifies three essential independent and collaborative modules, Decoy, Captor and Orchestrator. Based on the efficient architecture, a Software-Defined Networking (SDN) enabled honeypot system is designed, which supplies high programmability for technically sustaining the features for capturing high-quality data. A proof-of-concept system is implemented to validate its feasibility and effectiveness. The experimental results show the benefits by using the proposed architecture comparing to the previous honeypot solutions.

## I. Introduction

COMPUTER systems across the globe are faced with various security threats due to programming flaws and configuration errors. The consequences can affect individuals and organizations at critical levels from privacy exposures to financial losses [1]. Also, the hard to detect zero-day attacks are becoming increasingly numerous and sophisticated, e.g. there was an increase of 7 percent in the number of zero-day vulnerabilities recorded in 2017, and 27 percent of the 140 targeted attack groups that Symantec tracks have been known to use zero-day vulnerabilities at any point, which were shown by Symantec 2018's ISTR report [2]. Moreover, cyber threats are often accompanied with malicious motives: hack into government and private systems to cripple the military, force political progress, influence the financial markets, and damage the service sectors of economics [3]. Such cyber espionage, warfare and terrorism have provoked considerable alarm [4].

To reduce the risk, a variety of security measures are given, e.g. firewall, intrusion detection system (IDS), and intrusion prevention system (IPS). Unlike these tools mainly being used to prevent attacking, a honeypot is a specific security facility that aims to allow being attacked for the purpose of studying the creation of the hacker community by means of advertising/exposing its information systems resource to lure unauthorized and illicit access [5]. Besides the captured data quantity, the data quality is an even more significant aspect, which will greatly impact the attack investigation. With regard to the variety of attack types [6] and the complexity of attack scenarios [7], the multi-dimensional criteria based attack profiling and forensics are appealing [8]. For example, a typical DDoS attack takes a lot of compromised hosts to launch concurrent access to the target server that will result in a Denial of Service; a port-scan attack can use only one attack host but can involve a large number of victim hosts, however, it does little harm to the victims; a kind of buffer overflow attack can allow the adversary to break into the victim host by exploiting the vulnerabilities, which will be harmful because the attacker can access the victim's data and even can control the victim to carry out further attack; a malware can spread on the Internet and infect all the victims, which may cause a catastrophe globally, e.g. the WannaCry ransomware [9]. So the challenge is that a honeypot system should be capable of efficiently feeding the adversary with the appropriate resource depending on the attack type for the purpose of capturing high quality data.

However, most honeypots are only passively receiving attack data [10], and more precisely, they lack a sensibility for fully identifying and distinguishing the various attack data and scenarios. Despite a number of proposals providing some data control (honeywall [11], honeybrid gateway [12], honeyproxy [13]) and resource hierarchy (hybrid honeypots [14]–[16]) in order to address the problem, they are merely case-by-case and one-sided solutions. Furthermore, some countermeasures, e.g.

Manuscript received October 7, 2018; revised January 6, 2019; accepted January 11, 2019. This research is supported in part by Key Research and Development Program of China (No.2016YFB1000602), "the Key Laboratory of Space Astronomy and Technology, National Astronomical Observatories, Chinese Academy of Sciences, Beijing, 100012, China" National Natural Science Foundation of China (Nos. 61440057, 61272087, 61363019 and 61073008, 11690023). This research is also partially supported by the Spanish Ministry of Economy and Competitiveness in the context of GREDOS project (TEC2015-67834-R). (Corresponding author: Zhihui Du)

W. Fan is with the School of Computing, University of Kent, CT2 7NF, Canterbury, UK. (e-mail: w.fan@kent.ac.uk).

Z. Du is with the Department of Computer Science and Technology, Tsinghua University, 100084, Beijing, P.R.China. (e-mail: duzh@tsinghua.edu.cn).

M. Smith-Creasey is with the School of Mathematics, Computer Science and Engineering, City, University of London, EC1V 0HB, London, UK. (e-mail: max.smith-creasey@city.ac.uk).

D. Fernández is with the Department of Telematics Engineering, Technical University of Madrid, 28040, Madrid, Spain. (e-mail: david@dit.upm.es).

dynamic deployment [17], interesting traffic redirection [18], uninteresting traffic reduction [19] etc., are very useful to enhance the data quality. But current honeypots often either ignore providing countermeasures or barely offer simple ones. Besides, the manipulation on the attack flow and the honeypot resources should be stealthy and undetectable to the adversary, otherwise the further data capture will fail [20], [21].

Therefore, a type of efficient honeypot system, whereby the attack activities can be sensitively classified and processed in fine-grained ways, and then consumed by the unperceived honeypot resource with appropriate countermeasures is highly needed [22]. Nevertheless, the existing honeypot systems are not able to provide this kind of comprehensiveness, because there is a lack of architecture that can facilitate the all-round honeypot design. In [23], the authors identified the two kernel honeypot elements: decoy and captor, which can compose the honeypot infrastructure resource essentially, but the existing honeypot architectures often pay less attention on the captor, which greatly restricts the possibility of enhancing the high-quality data capture (see the following case 1).

Take the fishing trap as a metaphor of honeypots:
- Case 1: the hook with bait can catch naive and greedy fish (like script kiddies), but probably will fail to capture sophisticated fish (like advanced hackers);
- Case 2: separate the bait from the hook, and put it into a net, which will be more covert, and also will have higher fish capture efficiency.

Once the honeypot infrastructure is divided, the architecture often needs an orchestrator to enable them to cooperate. So, this research work's objective is to propose an efficient honeypot architecture that can orchestrate the two essential infrastructures so as to enable the all-round honeypot system design to satisfy the sensibility, countermeasure and stealth for various requirements with the purpose of capturing high-quality attack data. The contributions of this paper can be summarized as follows:
- An efficient honeypot architecture, namely HoneyDOC, consisting of three modules, i.e. Decoy, Orchestrator and Captor, is proposed to coordinate them so as to enable all-round design for the purpose of high-quality attack data capture.
- A SDN-enabled honeypot system is designed upon the proposed architecture. SDN's programmability and separated planes fully satisfies the requirement of facilitating the three significant features, i.e. sensibility, countermeasure and stealth, and makes the system extendable as well so that it is facile to develop new functions and integrate external third party components.
- A Proof-of-Concept system is implemented in terms of the SDN-enabled honeypot design, which is used to validate the proposed honeypot architecture, and also, some experiments are conducted for evaluating the features of the prototype.

The organization of this paper is as follows: Section 2 reviews related work; Section 3 presents the conceptual honeypot architecture; Section 4 proposes the SDN-enabled honeypot system architecture; Section 5 describes a proof-of-concept implementation of the honeypot system; Section 6 demonstrates the system evaluation and the experimental results; Section 7 makes a conclusion and suggests some future work.

## II. Related Work

This section reviews the conventional and SDN-enabled honeypot architectures. In the following content, the acronyms, LIH, MIH and HIH (refer to subsection III-A for definitions), stand for low-interaction, medium-interaction and high-interaction honeypot respectively.

### A. Conventional honeypot architectures

The Honeynet Project proposed a series of physical honeynet architectures [24] including Gen I, II, III, which were widely used by organizations. For example, Georgia Tech applied the Gen I honeynet architecture to manage compromised computers across the campus networks [25]. Thereafter, the virtualization technology was introduced to facilitate the virtual honeynet deployment, which makes one physical host running multiple guests [26]–[28]. However, this traditional honeynet architecture lacks the capability of large-scale deployment. In 2006, a hybrid honeypot framework was proposed [14], which integrated the famous low-interaction virtual honeypot framework Honeyd [29] with the Gen III honeynet architecture for improving IDSs to protect local production networks. Meanwhile, a number of hybrid honeypots emerged [15], [18], [30] since they can collect datasets on both of detailed attack processes and large network space coverage.

Among these hybrid honeypot architectures, the traffic redirection mechanism is aimed to connect the frontends and the backends. It is used to filter and redirect the interesting traffic into the HIH for in-depth analysis. The hybrid honeypot framework [14] simply used the Honeyd's built-in non-transparent proxy to redirect the traffic into HIHs. Due to the fact that this approach lacks a traffic filtering mechanism, the backends can be flooded by invalid data easily. Also, the non-transparent proxy approach has the identical-fingerprint problem since the frontends and backends assigned with different IP addresses. Some other hybrid honeypots [30], [31] used GRE tunnel to redirect the traffic, but the identical-fingerprint problem between the frontends and backends was still unsolved, and all traffic was merely treated by the frontend with two coarse-grained modes: discard or forward. Instead, in [12], [18], the TCP connection replay approach was applied for facilitating the connection migration from LIHs to HIHs. In particular, the transparent Honeybrid proxy/gateway was proposed in [12], where the author used the libnetfilter_queue [32] to process packets, so that security researchers can gain insight into the technical detail. However, they didn't address the identical-fingerprint problem.

Later, some solutions were presented to address the identical-fingerprint problem based on the transparent Honeybrid gateway. Lengyel et al. proposed a hybrid honeynets architecture namely VMI-Honeymon [33]. It used separate network bridges to isolate the original VM and its clones that retains the identical MAC and IP address. The authors believed this method can avoid the MAC and IP collision, even though the clones assigned with the same fingerprints are placed on the same network. Fan et al. [34], [35] proposed a dynamic hybrid honeypot system intending to address the identical-fingerprint problem, however this needs the honeypots to be frequently switched on and off.

B. SDN-enabled honeypot architectures

Software defined networking (SDN) aims to separate the system that determines the direction of traffic (control plane) from the underlying systems that forward traffic to the selected destination (data plane). Hence, the capability of flow control is the innate advantage of SDN. The programmable SDN-based network can allow the system administrator to dynamically configure the data plane according to the requirements [36]. The SDN technology are already widely used in the field of network security of distributed systems [37]–[39].

In recent years, several SDN-enabled honeypot systems have been proposed. HogMap [40] is a collaborative honeypot system essentially. It adopts SDN technology to simplify marketplace coordination across different domains to participate in HogMap, various providers with diverse network architectures only need to be equipped with an OpenFlow switch and the HogMap-certified SDN applications, which enable the provider to participate in various services and perform just-in-time actions to forward traffic without manual configuration. HogMap uses a packet replay based session migration mechanism. But it did not describe how to provide stealth, i.e. how to solve the identical-fingerprint problem.

HoneyMix [41] is another interesting SDN-based intelligent honeynet, which used to simultaneously establish multiple connections with a set of honeypots and select the most desirable connection to inspire attackers to remain connected. The confusion is whether the honeypots containing the same services use the the identical fingerprint. Unless the system addressed this issue, or else it may fail to pipe the connection between the honeypot and the switch to the one between the switch and the attacker.

HoneyProxy [13] proposed the essential component used in HoneyMix. The proxy module distributes the requests and selects the most appropriate response for the attacker to interact with. In order to deliver malicious traffic to relevant honeypots and select the most appropriate reply from multiple responses of the honeypots, the authors designed three modes: Transparent Mode (T-Mode) to forward the scanning or login attempts to an IDS equiped LIH; Multicast Mode (M-Mode) to delivery the payload packet to all associated honeypots and determine the best reply responding the attacker in order to counteract fingerprinting attacks; Relay Mode (R-Mode) to maintain an exclusive connection between the attacker and the HIH. This approach can effectively manage the connection and provide appropriate reply. However, because HoneyProxy represents the whole honeynet as a blackbox that runs many vulnerable services, the major drawback is that the HoneyProxy is a non-transparent proxy hiding all its inner individual small honeypots other than exposes them directly. One system involving many vulnerable services probably will cause the adversary's suspension, thereby it disobeys the principle of wide data capture and stealthy data control.

In [42], an intelligent honeynet architecture based on the SDS framework [43] is proposed, enabling flexible deployment and dynamic provisioning over Network Function Virtualization Infrastructure (NFVI). This paper focuses on migrating resources according to the workloads of each honeypot and power off unused modules, which is a kind of countermeasure that increases the cost-efficiency of the honeypot resource. However, that is not the essential issue in honeypot research context. The design of the traffic forwarding component is unclear: if it is non-transparent, then it has the similar problem with HoneyProxy when dynamically dispatching the traffic; If it is transparent, then it lacks the way of providing stealthy traffic and resource migration.

III. HoneyDOC Architecture

The HoneyDOC decouples the Captor and the Decoy from the architectural point of view, and by using the Orchestrator to coordinate them, it can efficiently enable all-round honeypot design. We already noticed that the difference between a honeypot and a vulnerable system is that a honeypot must be trusted but a vulnerable system is untrusted. So, honeypot must have some security program to make it trusted. In [22], the author showed that the security level should be enhanced along with the interaction level in context of honeypot. The paper [23] separated the captor from the decoy, and the captor actually represents the security program in context of honeypot. However, during the past years of honeypot development, the captor has not received enough attention, and the proposals were often fat decoy and thin captor solutions, and in most cases, they were not decoupled and the importance of the captor is often overlooked (see case 1 of the metaphor in the Introduction section). But the fact is that the captor is playing an increasingly important role in (high-quality) data capture. Previously, there was no honeypot architecture proposal endorsing the captor to have a peer status with the decoy, which inevitably resulted in the weak or limited captor functionality in the honeypot.

Therefore, the decoupled Decoy and Captor can unleash the power of the captor for serving the high-quality data capture, meanwhile, the decoupling brings benefits to the Decoy as well, since it can be more flexible and diverse once it leaves the constraints of the Captor. Also, the decoupling is the basis for flexibly combining the Decoy and Captor to perform a powerful honeypot system (see case 2 of the metaphor in the Introduction section). Now that the Captor and Decoy are decoupled, on one hand, they can be developed respectively and their capability can be updated independently. On the other hand, they can be combined together in different ways so as to carry out more powerful functionalities. However, to facilitate the combination is not an easy task, which involves job dispatch, function interaction, etc. so then a new significant module named Orchestrator is highly needed for the purpose of coordinating the two modules to run as a whole honeypot system.

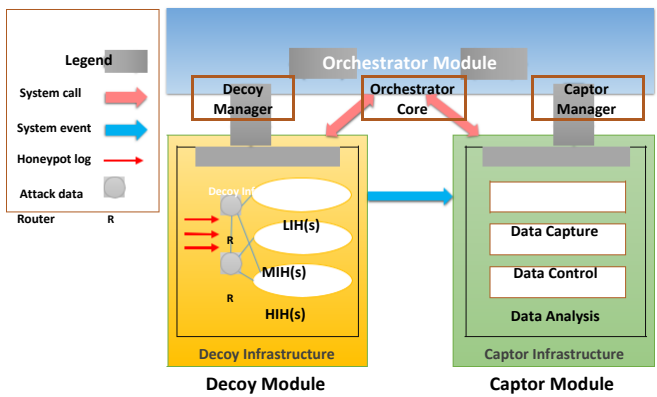

Fig. 1. An overview of the HoneyDOC architecture

So, we propose a novel honeypot system architecture (see Fig. 1) that decouples the Decoy and Captor into two separate modules, where they will have an equally important status, and they are coordinated by the Orchestrator module so as to work them together in a more efficient way. In the following subsections, each module is described in detail respectively.

A. The Decoy module

The Decoy module is responsible for provisioning and deploying the honeypots/honeynet over the decoy infrastructure. The decoy infrastructure should be able to host the deployment of different types of decoy. Also, it needs to expose its API that the Orchestrator module can call. As the conventional honeypots often do not decouple the decoy and the captor, we use "decoy" and "honeypot" interchangeably in this subsection.

At present, a great number of dedicated honeypot softwares have been developed [10], which can be roughly classified into three categories in terms of the interaction levels. A low-interaction honeypot (LIH) is a program that emulates the protocols of an operating system (OS), but with a limited subset of the full functionality. For example, Honeyd [29] can emulate multiple decoys simultaneously to monitor the unauthorized traffic. A medium-interaction honeypot (MIH) can provide much more interaction. It can often emulate a variety of vulnerable services based on the TCP/IP network stacks that are implemented and managed by the underlying OS. However, the MIHs, e.g. Dionaea [44], only emulate well-known vulnerabilities and capture malicious traffic accessing to them. A genuine computer system running as a honeypot is called high-interaction honeypot (HIH), since it can provide a fully functional OS for attacking. Using HIHs, security researchers can capture not only the network activity, but also the system activity. The limitation of HIHs is the resource consumption for large-scale deployment. Though the dedicated honeypots are effective for meeting single criterion, they are difficult to satisfy multiple criteria, since they are either expensive in scalability (HIHs) or hard in collecting the detailed attacking data (MIHs and LIHs).

Therefore, our Decoy module aims to host three types of decoys. As multiple decoys can be deployed as a honeynet by following a certain network topology, this module should also allow to deploy arbitrary honeynet. The attack data flow goes through the decoy infrastructure, and the attack data should be captured so as to yield the honeypot log which then will be transferred to the Captor module. Some attack data will cause system event that will be sent to the Orchestrator module.

B. The Captor module

The Captor module is in charge of providing the functionalities that can be applied to the attack data. This module comprises three basic submodules: data capture, data control, and data analysis. Nevertheless, the goal is to have an extendable module, so the submodule is not limited to the above three ones.

1) Data Capture: The purpose of data capture is to log all the intrusion events and malicious behaviors for later investigation. It is a compulsory functionality, so any type of honeypot system must have this functionality. Three critical layers of Data Capture were identified: firewall logs (inbound and outbound connections), network traffic (every packet and its payload as it enters or leaves the honeypot), system activity (attacker keystroke, system call, modified files, etc.). The more data and the higher quality of the data which the honeypot can capture, the better the honeypot system is.

2) Data Control: The data control functionality is aimed to conduct the attack flow according to the honeypot's intention instead of the attacker's. So, it often needs to be stealthy and transparent to the attacker. The data control actually provides the countermeasures against the intrusion. For the inbound attack flow, it can discard the uninteresting data, forward the interesting data, and even redirect the most interesting data to dedicated decoy resource. For outbound attack flow, it is used to mitigate the risk that the adversary uses the compromised honeypot to attack other non-Honeypot systems. Any HIH must have this functionality. Also, it is necessary to minimize the attacker or malicious code chance of detecting it. The outgoing data can be simply dropped, but the challenge is how to set the threshold: the more you allow the attacker to do, the more you can learn; however, the more you allow the attacker to do, the more harm they can potentially produce. At present, there are several other solutions to control the outgoing traffic such as transparently redirect the outbound connection to another honeypot that emulates the target system.

3) *Data Analysis:* The entire purpose of data analysis is to analyze the collected data in order to get the information of the attack. For example, through data analysis the technique of attack and the adversaries motivation could be revealed. Thus, if the data cannot be analyzed, the value of honeypot system will be reduced. This submodule can employ and integrate the third party analysis service. In practice, some data analysis process, e.g. comprehensive forensics, could be performed by security experts or by automated programs.

### C. The Orchestrator module

The Orchestrator module takes the responsibility of coordinating the other two modules to work together efficiently. It consists of Decoy Manager (DM), Captor Manager (CM) and Orchestrator Core (OC).

The DM can actively call the decoy infrastructure API in order to provision and deploy the needed decoy or a network of decoys. So, if the decoy infrastructure is dynamic, the DM can request redeploying the decoy(s) on demand even in real-time. This will be useful to facilitate a type of countermeasure - dynamic deployment.

On the other hand, the CM is used to integrate the captor infrastructure by calling its API. The CM can call the captor to process the system event as well as the honeypot log. Different captors have different functionality, so the CM is responsible to execute the right call based on the decision from the OC.

The OC is the kernel of the Orchestrator module. Firstly, all the system events will be handled by the OC. The system events from the decoy infrastructure and the captor infrastructure can be directly sent to the OC, which can dispatch the inbound events to the appropriate component for process, and it can also generate outbound events as reaction or countermeasure. Secondly, it will conduct the interaction between the DM and CM, scheduling the call to their APIs in order to perform the needed operations.

One note is that the honeypot log sending from the Decoy module to the Captor module should not rely on the Orchestrator module. That means even there is no orchestrator, the basic honeypot function - data capture - is still working, but then the honeypot system's functionality will be very simple. So, without the orchestrator, is will be very hard to apply the captors to the attack data in order to improve the data quality.

## IV. SDN-enabled HoneyDOC Design

In this section, we first define the three advanced honeypot features as our requirements, then we will introduce the SDN to sustain our all-round honeypot design, and afterwards we will present the SDN-enabled HoneyDOC design and explain how it facilitates these three features. Finally, we will highlight the novelty of our work by comparing it with the literature in terms of the contribution into the three features.

### A. Three Key Features

As aforementioned, the features named sensibility, countermeasure and stealth are vitally important to the honeypot for the purpose of catching high quality attack data. In this part, we provide their definitions in the context of honeypot system as follows:

- Sensibility: the honeypot system has a keen consciousness to detect various attacks, and also is able to classify and process the attack data in fine-grained ways. It represents an advanced detection capability instead of merely alerting malicious activity.
- Countermeasure: the honeypot system can provide response to the attack with the purpose of capturing high-quality attack data instead of being a victim.
- Stealth: the honeypot system's functions and operations against the attack behaviours keep in stealthy in order to prevent them being detected by the adversary. This feature can guarantee the effectiveness of honeypot.

According to the definitions, we can say that Sensibility and Countermeasure are two functional features, while Stealth is a non-functional feature. Nevertheless, the stealth feature must be reflected in the former two functional features, otherwise the other two will be worthless. So, the importance of Stealth is not more than enough. In this work, we will mainly address the problem related to the stealth requirement for honeypot system.

### B. The SDN Technology for Honeypot

A traditional SDN architecture has three planes: application, control and data plane. The application plane is separated from the control plane by the SDN Northbound Interfaces (NBI). They are not so decoupled like the control plane and the data plane, because the applications are often developed upon the specific SDN controller software. In the data plane, the network elements (NE), e.g. switches and routers, can be linked by arbitrary topology, and the end points (e.g. PC, laptop, etc.) can be integrated into the network by linking any NE. All the conversations over the data plane are controlled by the programmable control plane through the Control-Data-Plane Interfaces (CDPI). So, SDN has the advantages of ease of management, efficient flow control and extendable integration.

Thereby, upon the current technique background, SDN is a very competitive sustaining technology to facilitate HoneyDOC system design. The SDN's features, especially the decoupling and programmability, can satisfy the requirement of sensibility, countermeasure and stealth.

Indeed, the SDN technology can tailor HoneyDOC system's one-stop solution. The data plane takes the responsibility of directing the traffic to the heterogeneous decoys with user-defined network topologies deployed by the system. The control plane can use a specific SDN controller software to provide the network services for supervising and managing the data communication over the data plane. The application plane can include various captors upon the SDN controller's framework. In contrast, HoneyDOC unleashes the power of SDN's programmability for enforcing the vitally significant function -data control- in the context of honeypots as well. Table I summarizes the mapping of the HoneyDOC modules to SDN planes for the purpose of clarifying how this architecture can be cohesively represented in SDN.

TABLE I
Mapping of HoneyDOC architecture to SDN Framework

| SDN Framework | HoneyDOC |
|---|---|
| Application Plane | Captor |
| Control Plane | Orchestrator |
| Data Plane | Decoy |

### C. SDN-enabled HoneyDOC System

The SDN-enabled HoneyDOC system includes an application (i.e. Captor) plane, a control (i.e. Orchestrator) plane and an infrastructure/data (i.e. Decoy) plane. The captor plane consists of various captors, the control plane needs to adopt a specific SDN controller software and the decoy plane deploys heterogeneous honeynets. The captors are developed/integrated based on the adopted SDN controller's APIs, i.e., the Network Data Control Application is designed, which consists of a Decision Engine (DE) and a Redirection Engine (RE): the DE is in charge of the action of processing the traffic and the RE enforces the action. Figure 2 shows the SDN-enabled HoneyDOC system design.

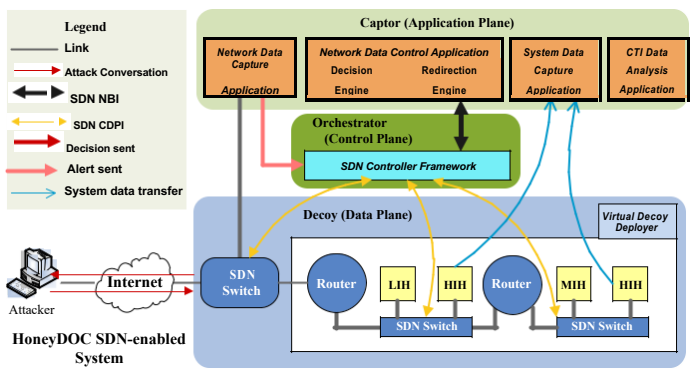

Fig. 2. An overview of HoneyDOC SDN-enabled System Desgin

This system is set up to receive traffic destined to decoys. The incoming traffic is firstly classified, the uninteresting traffic will be filtered or forwarded to LIHs, and the remainder is treated as interesting data being redirected to appropriate HIHs by the out ports of the SDN switch. Any malicious behavior in HIHs will be captured, and the outbound traffic from them will be controlled to avoid attacking the non-honeypot systems.

The network data capture and control applications work collaboratively to facilitate the traffic classification and redirection, which are used to process and control the network data flow forwarding to the decoys. That is sensitivity relevant, so will be described in more detail in subsection IV-C1. The virtual decoy deployer is in charge of configuring, creating and managing heterogeneous decoys for data capture. It should be flexible enough to deploy different single dedicated decoys in terms of the interaction levels, and also should be able to handle a complete honeynet including various decoys. Furthermore, the system data capture application is responsible for catching the system activity in the HIHs. Also, the cyber threat information (CTI) data analysis application is aimed to investigate the logged CTI data in order to reveal further cyber threats, whereby appropriate reaction can be carried out in advance. These countermeasure related components will be described in more detail in subsection IV-C2. Basically, all the components should work in stealth against adversary's suspicion. In this paper, we mainly focus on the stealthy TCP connection migration for redirecting the attack flow, which will be described in subsection IV-C3.

*1) Sensibility:* Sensibility can greatly increase the efficiency of data capture by the measure of classifying and filtering the inbound network traffic.

*a) Multiple classification criteria:* The Network Data Capture Application works with the DE to provide a customizable traffic classification approach, which allows the user to set arbitrary rules and associate actions (i.e. Drop, Forward, or Redirect). There are several ways to make traffic classification: signature-based (i.e. payload-based) and source-destination based (i.e. addresses-based). Due to the fact that the answer to which traffic is worth being investigated is subjective and depends on the security researcher's intention, the system should allow the user to customize the traffic classification policy. In order to support multiple traffic classification criteria, we apply a rule-based way to provide the customizable traffic classification. The user can set the "action" field in the rule for processing the matched traffic. We consider integrating an NIDS to facilitate the Network Data Capture Application to evaluate the traffic with the purpose of sending the alert message to the DE so as to make corresponding decision on the traffic.

*b) Fine-grained process method against the classified data:* The RE will carry out according to the "action" field of the alert message. If that is "DROP", the RE will discard the traffic, ending the connection. If the alert message indicates that the traffic has to be forwarded to either an MIH or an HIH, the RE will continue processing the flow, selecting the corresponding SDN switch's (FCF, see detail in Fig. 3) out port that links the the target honeypot and forwarding the packet to it. Meanwhile, the required SDN flow entries will be installed in the corresponding SDN (SPF, see detail in Fig. 3) to do the TCP sequence number synchronization. Consequently, once the TCP connection has been transferred to the target honeypot, the NIDS does not need to inspect the subsequent packets of that connection any more.

2) *Countermeasure:* Countermeasure provides the response against the attack behavior with the purpose of enhancing the efficiency of data capture. In [37], several countermeasures against intrusion were presented. Likewise, regarding to the honeynet scenario, the countermeasures can be divided into three categories: attack flow control, decoy dynamic deployment, data analysis and vulnerability fix.

   a) *Attack flow control:* This type of countermeasure is the main one with regard to the Data Control described in subsection III-B2. The traffic can be blocked, discarded, redirected or isolated, upon SDN technique, all these flow manipulations can be facilitated by the Network Data Control Application aforementioned.

   b) *Decoy dynamic deployment:* Decoy dynamic deployment is with regard to the decoy timely revolution. Apparently, it is a tedious task to configure and deploy a honeynet manually. The virtual decoy deployer is designed to dynamically deploy and manage the honeynet. It supports some concrete countermeasures. For instance, dynamically deploy a decoy to receive the migrated interesting traffic; emulate the non-honeypot system to contain the back-scatter/outbound traffic; reconfigure the decoy's fingerprints and the honeynet topology to reduce the possibility of being detected.

   c) *Vulnerability fix:* Vulnerability fix rides on the CTI data analysis, and they often result in a joint countermeasure. CTI data analysis is the concrete reflection to the Data Analysis (subsection III-B3). The HoneyDOC system enables the third party analysis framework to be integrated, e.g. the user can integrate the C3ISP framework's gateway for the purpose of sharing the CTI data to consume the analytic services [45]. Consequently, the system needs to fix the vulnerabilities, which can be achieved by patching the OS and software, or setting the flow filtering rules.

3) *Stealth:* Stealth aims to deceive the adversary to behave as in its self-righteous network environment so that it can show off its attacking ability, which then can be fully observed by the honeypot system. Basically, both the Decoy and Captor should be stealthy: (1) Decoy's stealth relies on its fidelity, apparently, the virtual honeypot with lower interaction or worse performance often has the higher probability to be detected, however, the rapid improvement of virtualization techniques can be used to address this problem efficiently; (2) Captor's stealth issue often comes up with the data capture and data control functions. Concerning the system data captor, that must be careful whether it is able to catch the activity but is not detectable. It is a complex issue (referring to subsection V-D for some detail) which this paper does not focus on.

On the other hand, for the data control, in particular, the network traffic redirection needs to be stealthy and transparent to the adversary, which is a significant issue. We will present a solution based on the SDN to address this problem in the following content.

As stated, the traditional traffic redirection solutions have the identical-fingerprint problem, which means that the server receiving the transferred flow has a different fingerprints, i.e. MAC and IP addresses, from the previous server that established the original connection. In common network service environments, that is not a problem. However, in a honeynet scenario, the different fingerprints will lead to the attacker's suspicion. Furthermore, concerning the traffic redirection, the header information of three protocols -Ethernet, IP and TCP- needs to be handled. Ethernet and IP protocols are stateless so that only the destination fields of the packets' headers need to be modified. Regarding the transport layer protocols, UDP is stateless as well, however, TCP is stateful whereby multiple fields need to be updated for migrating the connection. The TCP mechanism includes two sequence (Seq) numbers randomly generated by the source endpoint and the destination endpoint separately, then acknowledged and incremented by each end-point's network stack to guarantee that all the packets are transmitted correctly. So, when transferring a TCP connection, the endpoint receiving the migrated connection normally will create a different Seq from the the original endpoint's. Thereby, to facilitate the transparent TCP connection handover, the main issue is to synchronize the Seq and Ack values related to the sequence numbers chosen by the different endpoints.

Hence, two technical challenges have to be addressed to facilitate the stealthy traffic redirection: 1) identical-fingerprints of different decoys; 2) sequence number synchronization for connection transfer. For solving the first problem, we will use the distinct out ports of the SDN switch to identify the end-points with the identical IPv4 and MAC addresses. For addressing the second problem, we use the TCP replay approach [21] to transfer the connection, which updates the Seq number on-the-fly by the SDN controller. Figure 3 illustrates the design of the approach, which has the advantage of reducing the computational burden of the controller. In order to make the approach feasible, two different functions have to be performed: 1) the Flow Classifying Forwarder (FCF) is used to identify and isolate the end-points; 2) the Session Processing Forwarder (SPF) is responsible for enforcing synchronization. Each honeypot designed to receive the redirected connection must be equipped a SPF in front of it for the purpose of adapting the TCP sequence numbers.

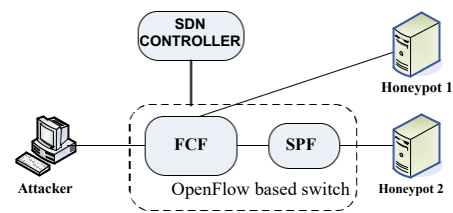

Fig. 3. The approach for synchronizing Seq and Ack numbers

Based on this design, two SDN-based TCP redirection mechanisms are proposed: the first mechanism uses the controller as the frontend to establish the first TCP connection with the attacker, while the second mechanism applies a VM as the frontend to respond the TCP connection request. They are graphically illustrated in Figure 4 and Figure 5 separately. These two mechanisms have the similar working phases of traffic redirection such as the diagrams show. We take the mechanism 1 as the example to describe the three phases:

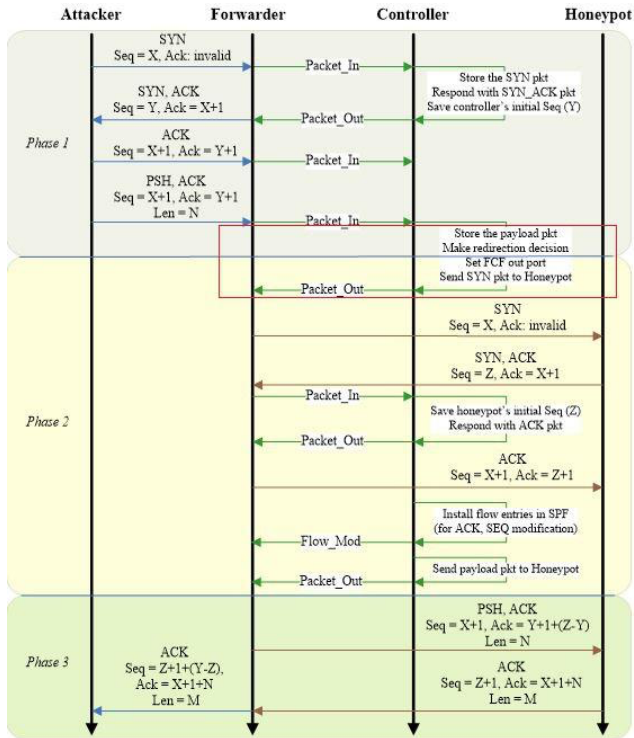

Fig. 4. TCP Migration Mechanism 1: using the controller as the frontend.

1) Phase 1: The session between the attacker and the controller is established. The attacker sends the TCP connection request to the honeypot. The controller answers the request on behalf of the honeypot.
2) Phase 2: If the controller decides (see subsection 4.1.2) to redirect the connection, it selects a suitable honeypot based on the decision process and instructs the FCF to send the session segments to the honeypot port. The TCP session is transferred from the controller to the honeypot by replaying the initial session segments. The flow entries with synchronization functions are installed into the SPF.
3) Phase 3: The TCP session has been transferred to honeypot and Seq and Ack numbers are synchronized. After phase 2 changes, the system is configured to send the session traffic directly

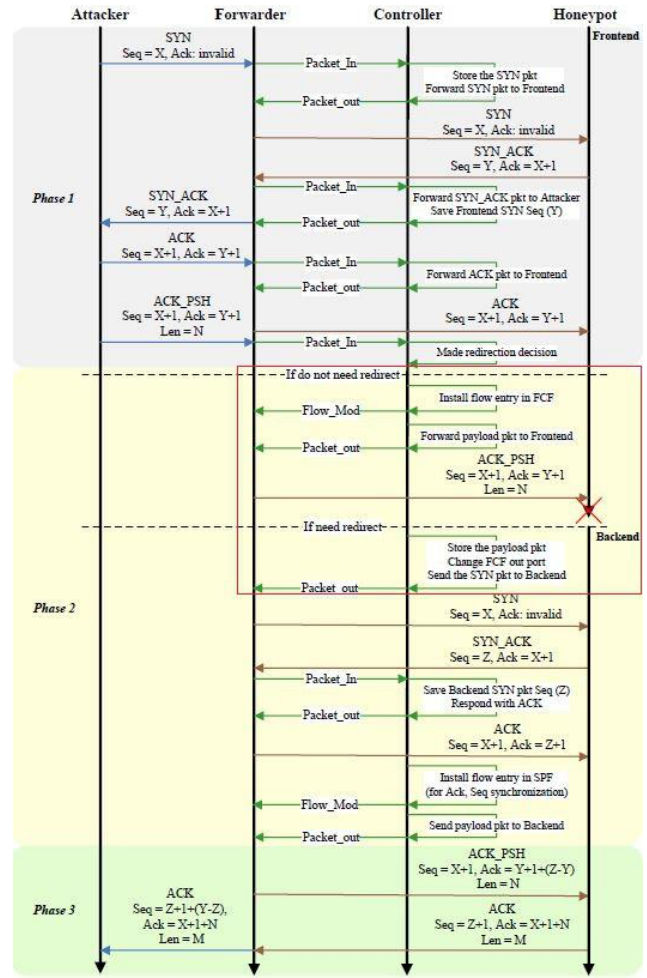

Fig. 5. TCP Migration Mechanism 2: using the VM as the frontend.

between the attacker and the honeypot, having the SPF doing the synchronization.

We use the red boxes to highlight the important steps and difference between the two mechanisms. Mechanism 1 always needs to store the payload pkt and create a new TCP connection to the honeypot; Mechanism 2 acts relying on the decision, if it does not need redirection, the payload pkt will forward to the original VM, but if it needs redirection, the payload pkt will be stored, and a new TCP connection to the new VM will be established. The technical detail of the SDN-based TCP connection handover as well as the sequence number synchronization based on the second mechanism has been presented in [21]. Besides, the process to establish connections that we have discussed above focuses on the inbound traffic, but can also be applied to the outbound traffic, deciding to redirect or discard honeypot outbound connections as part of the containment mechanisms of the system.

D. Literature Comparison

In this subsection, we summarise the significant advancement of our work by comparing it with the literature work in terms of the three features proposed in this work. Table II presents the comparison, which illustrates the improvement created by HoneyDOC over the Sensibility, Countermeasure and Stealth.

TABLE II
Literature Comparison in terms of the three features

| Work | Sensibility | Countermeasure | Stealth |
| --- | --- | --- | --- |
| HoneyBrid | Low | Low | Low |
| VMI-Honeymon | Low | Low | Medium |
| HogMap | Low | Medium | Medium |
| HoneyMix | Low | Medium | Medium |
| HoneyProxy | Medium | Medium | Low |
| SDS-based Honeynet | Medium | Medium | Low |
| HoneyDOC | High | High | High |

## V. Proof-of-Concept Implementation

In this section, a proof-of-concept system is presented in order to conduct the validation experiments. We implement the whole system on one physical machine as Figure 6 shows (it uses the traffic redirection mechanism 1, to enforce mechanism 2, the SPF in front of the LIH/MIH is removed). All components of the system are implemented by using specialized tools (see following subsections), which are not limited to those used in this prototype. The workflow can be summarized as follows:

1) The user prepares the configuration of the honeypot scenario based on TIHDL [46], configures the NIDS Snort[1] for setting the traffic classification rules, and configures the SDN controller Ryu[2].
2) The Honeyvers is invoked to create the honeypot scenario according to the configuration.
3) The Deep Security Manager[3] is turned on to monitor the HIHs and log the CTI data.

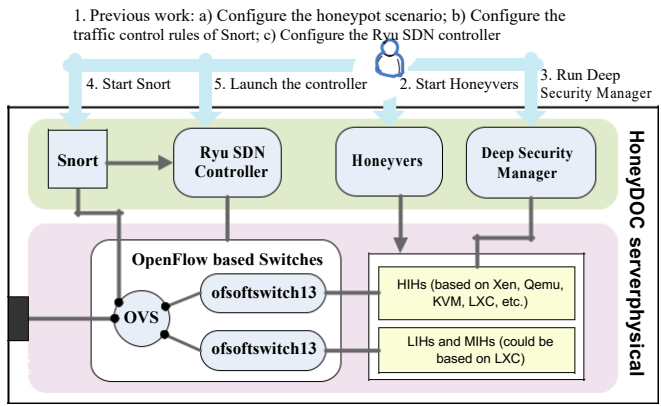

Fig. 6. An implementation of proof-of-concept by one physical machine

[1] https://www.snort.org/
[2] https://osrg.github.io/ryu/
[3] https://help.deepsecurity.trendmicro.com/software.html

4) Snort starts to listen on the corresponding interface linking the out port of the Open vSwitch (OVS).
5) The Ryu controller launches to receive connections and control the network flow.

### A. Rule-based NIDS

The well-known open-source NIDS Snort is used to facilitate the Network Data Capture Application. Snort is a rule-based NIDS aimed at detecting malicious traffic patterns that match the well-known signatures. We use the rule format of Snort as the basis to set our own traffic classification rules. A typical "alert" rule format of snort can be shown as follows:

> alert protocol source-ip source-port → destination-ip destination-port (msg: "alert message"; sid: an integer; priority: an integer; content: "malicious pattern";)

The text using bold font are the key words, and the text using italics font needs to be replaced by the values. The signature includes not only the IP header information but also malicious payload pattern. Snort checks on both of them in the captured packet. If there is more than one rule, we can set the "priority" field for each rule and use the "sid" field to set the alert ordering.

Therefore, based on the "alert" rule format of Snort, we can set reaction message into the "msg" field. For our own rule, we define three actions: DROP, MIH, HIH. "DROP" refers to discard the packet, while "MIH" and "HIH" indicates the packet forwarding destination. To implement the traffic classification approach, we apply the Snort rules by two steps to carry out both the source-destination based and content based traffic filtering. In the first step, Ryu controller will read and parse the Snort rules, translate them into SDN flow entries, and install them into the main OVS switch during the system initialization phase. This step aims to set a data reduction measure to improve the data capture efficiency. Only the rules associating with a "DROP" action and having blank "content" field will be translated into "drop" flow entries. Consequently, any traffic that matches these flow entries will be efficiently discarded in the data plane to avoid them reaching the controller. The other rules will be translated into "allow" flow entries, and the traffic that matches them will be forwarded as a PacketIn event to the Ryu controller for processing. In the second step, the Ryu controller will cooperate with Snort to perform the content-based traffic classification (see Figure 7).

Snort works in the NIDS mode so that it can utilize its intrusion detection and alert raising functionality. We use the snortlib, which can be accessed from the Ryu SDN framework, to integrate the Snort alert function into the Network Data Capture Application. The controller application will send the first payload packet of each connection to the Snort port. Snort thereafter inspects the payload packet, raises an alert, and sends the corresponding action message as the alert event back to the application through a Unix socket.

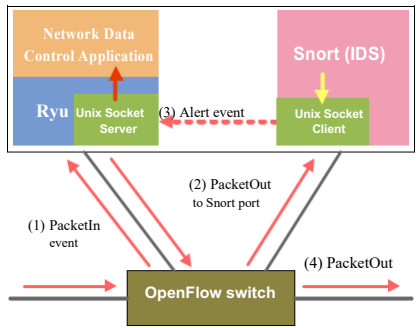

Fig. 7. Integrating Snort into Ryu framework for decision making

### B. SDN controller and switches

The Network Data Control Application is the most important component in the HoneyDOC SDN-enabled system. An SDN controller can satisfy both the system requirements and the ease of development is the needed one for proof-of-concept. The Ryu SDN framework is an open-source software and supports OpenFlow 1.3, which is well documented and provides easy to learn interfaces for application development. Thereby, the Ryu based prototype was implemented. For the traffic redirection mechanism 1, the controller application only provides the function of simulating open ports, but lacks the simulation of OS and fake service fingerprints, which would need more effort on developments. The adversary using scanning tools such as Nmap[4] will not be able to guess the exact OS, as we demonstrate in the experiment section.

The OpenFlow based switches are the most important facilities in the data plane. As Figure 3 shown, the FCF could be a physical or a software switch that supports OpenFlow. OVS is chosen as the main switch of the system because it is a pure software implementation that runs over standard Linux systems and it is implemented inside the Linux kernel, which makes it efficient when processing packets. Indeed, any Openflow compliant switch can stand in for OVS. On the other hand, due to the specific requirement of implementing the Seq and Ack numbers synchronization function in the SPF, we have adopted a different software switch, Ofsoftswitch13[5], since it is implemented in the user space that results in the ease of modifying the implementation code. Additionally, Ofsoftswitch13 is open source and can be freely forked from the Github project. We have modified the Ofsoftswitch13 to include the additional Seq and Ack synchronization functions. The synchronization functions are achieved through adding a new action based on the SET_FIELD defined by the OpenFlow 1.3 standard. These two actions are SET_TCP_ACK_DIFF and SET_TCP_SEQ_DIFF that are used to modify the Ack and Seq numbers respectively.

[4] https://nmap.org/
[5] https://github.com/CPqD/ofsoftswitch13

### C. Virtualized deployment tools

Owing to the advantages of easy management and resource efficiency, virtual honeypots have been widely used instead of physical ones [47]. The use of the virtual decoy deployer results in the need of integrating virtualized tools to deploy virtual decoys. Every specialized tool has its own specification to define the topology, addresses, types of systems, etc. Most MIHs can only emulate stand-alone decoy, e.g. Dionaea [44] and Amun [48], which often rely on some scalable deployment tool. Also, a virtual HIH decoy needs guest virtual machine as the carrier to contain it.

In order to hide the technical dependent complexity of the underlying tools, a generic virtual decoy deployer called Honeyvers [49] with a technical independent honeynet description language (TIHDL) [46] was proposed. It offers the versatility to manage heterogeneous decoys by the means of integrating the corresponding virtualized deployment tools, whereby the decoy states described by TIHDL can be translated to the corresponding states sustained by the concrete deployment tools.

### D. System activity monitoring and capture tool

The traditional kernel modification based tool Sebek/Qebek [50] had been proved easy to be detected. In contrast, the "out-of-the-box" method based system behavior monitoring tools often have the platform dependency and function limitation [23].

In this implementation, the Deep Security Manager (DSM) of TrendMicro is applied. The DSM is an off-the-shelf production supporting all popular platforms (i.e. Windows, Linux, etc.). The DSM includes six security modules (firewall, anti-malware, web reputation, intrusion prevention, integrity monitoring, log inspection) to fully monitor and capture the activities occurred in the target HIH. The DSM agent needs to be installed into the HIH. The malicious behavior occurred in the HIH will trigger the DSM agent, which thereafter can create CTI data (log) and automatically send back the data to DSM server. Further, the CTI data sharing for collaborative analysis can refer to this paper [45].

## VI. Experiments

As the prototype has been developed, we deployed a testing scenario to validate our proposal (see Figure 8, which is in regard to the traffic redirection mechanism 2). It includes one internal network 10.1.1.1/24 where the honeynet is deployed and one external network 10.1.0.1/24 where the attacker locates. A router stands in the middle of these two networks. The same IP and MAC addresses were assigned for both of the MIH and HIH so that they have the identical fingerprint. In the following content, we will present the validations of sensibility, countermeasure and stealth.

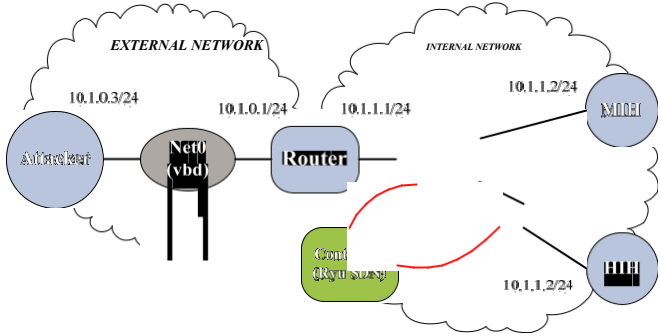

Fig. 8. The testing scenario: the dotted lines indicate the communication between the SDN controller and switches

### A. Validation of sensibility

The user can customize arbitrary detection rules with specific actions for classifying the traffic. For instance:

```
alert tcp any any → any 21 (msg:"MIH"; sid:1000002; priority:2;)
...
alert tcp any any → any 25 (msg:"HIH"; sid:1000005; priority:2;)
...
alert tcp any any → any any (msg:"DROP"; sid:1000008; priority:0;)
```

As the system has several open ports, any traffic requires to access to these ports will be processed by the controller, and then the controller will forward the traffic to the destination according to the Snort alert message. Any other uninteresting traffic (that matches the "drop" rule) will be filtered by the OVS, which can prevent the controller from traffic congestion. After the initialization of the controller application, the corresponding flow entries in the OVS are created as follows:

```
OFPST_FLOW reply (OF1.3) (xid=0x2):
cookie=0x0, duration=90.798s, table=0,
n_packets=0, n_bytes=0, priority=2,tcp,tp_dst=21
actions=CONTROLLER:65535 ...

cookie=0x0, duration=90.798s, table=0,
n_packets=0, n_bytes=0, priority=2,tcp,tp_dst=25
actions=CONTROLLER:65535 ...

cookie=0x0, duration=90.798s, table=0,
n_packets=0, n_bytes=0, priority=0,tcp actions=drop
```

So then after starting the system with a rule setup as aforementioned, only the requests associating to the open ports (e.g. port 21 and 25 shown above) can pass through the corresponding entries, the rest will match the last drop entry and be denied.

### B. Validation of countermeasure and stealth

This part will validate the traffic redirection that covers both the countermeasure and stealth. The proposed TCP connection migration aims at transferring the interesting traffic from the LIH/MIH (frontend) to the HIH (backend), which must be stealthy to deceive the adversary keep attacking. Figure 9 shows a flow control use case, where if the MIH and HIH uses different IP addresses, the attacker can easily realise the redirection after compromising the HIH by typing the simple "ifconfig" command.

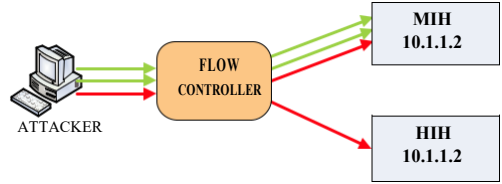

Fig. 9. Traffic redirection use case: the red dotted line indicates that the intended connection is redirected

For validating the TCP connection handover mechanism, we use the SSH to establish the TCP connection and apply Wireshark to observe the connection changes occurred between the frontend and the backend. Figure 10 shows the observation.

Fig. 10. The Wireshark flow graphs of the SSH redirection testing

The upper flow graph shows the initial TCP connection established between the attacker and the frontend. Thereafter, the ACK_PSH segment from the attacker is stopped sending to the frontend, since the Snort makes an alert indicating the controller to redirect the traffic to the backend. Thus, the controller starts to replay the TCP three-way handshake as the bottom graph of Figure 10 shows. After finishing the new TCP establishment between the attacker and the backend, the remaining segments are exchanged fluently between them. Meanwhile, the segments retransmitted from the frontend to the attacker are dropped by the controller, and finally the old TCP connection is terminated. Here we should note that the time displayed by Wireshark is relative to the first packet it captures by the network interface, so the time has no the reference meaning. Once the attacking connection is redirected to a HIH, the attacker can check on the destination address, but no change will draw his attention.

### C. Performance tests

For the performance evaluation, we design a test based on the SMTP protocol to monitor the latency of the first push packets arriving at the honeypot under concurrent inbound connections. An SMTP server (Postfix) was installed in honeypots. An SMTP client script was installed on the remote attacker. The script consists of the following sequence of five SMTP commands:

```
HELO test \n
MAIL FROM: <test@test.test> \n
RCPT TO: <root@localhost> \n
DATA. \n
test. \n
```

The experiment consisted of running the automated SMTP client script at the rate of 10 connections per second. We recorded the duration for the first push packet of each connection arriving at the honeypot. The results under different scenarios are shown in Fig 11.

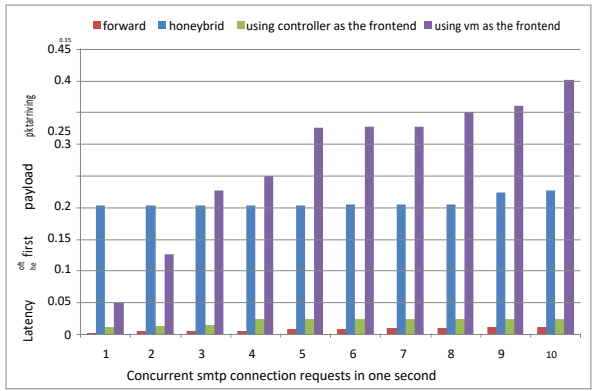

Fig. 11. Connection latency under different scenarios

The first push packet including payload arriving at the honeypot means the TCP connection between the attacker and honeypot has been established. So the timestamp of the first push packet arriving at the network interface of the honeypot can be used to calculate the duration for establishing TCP connection. The experimental results show that the connections processed by the Honeybrid gateway and the traffic redirection mechanism 2 can cause much more latency than the normal forward connections, while the connections processed by the traffic redirection mechanism 1 have short latency that close to the normal forward connections. Mechanism 1 uses the controller to play the role of frontend to answer the TCP connection request, which loses some fidelity and reality. Mechanism 2 uses the VM (a honeypot) as the frontend, it can respond to the attacker's request with real fingerprints. Hence, mechanism 1 has less latency, so it is more suitable for the case of capturing automatic attacks, while the mechanism 2 is more realistic, so it is more suitable for the case of capturing manual attacks.

Fig. 12 shows the packet I/O graph by using different mechanisms. The adopted interval is 100ms, so the diagram refers to the number of packets is processed by the honeypot in each 100ms. Within this interval, we can observe the packet I/O is more equal-distributed when using the traffic redirection mechanism 2 or never using any redirection. The packet I/O distribution of the traffic redirection mechanism 1 and Honeybrid is quite similar.

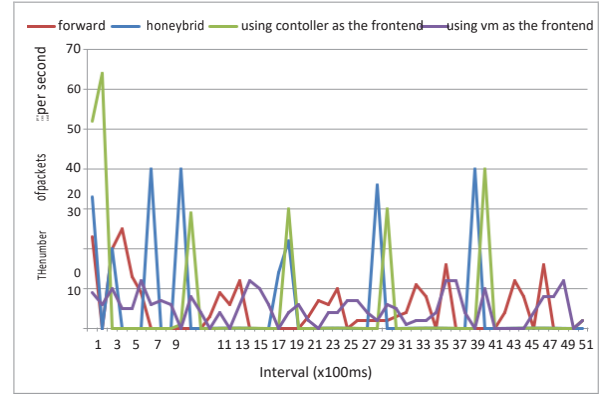

Fig. 12. Packet I/O graph of honeypot under different mechanisms

These two experimental results show that the added traffic control mechanisms provide more sophisticated functions than the state-of-the-art but do not cause performance loss.

### D. Live attack capture

In order to capture real attack data, we deployed the system including multiple decoys with two identical fingerprints in a low security production network of Universidad Politécnica de Madrid (UPM) (see Figure 13). On one hand, owing to the security requirements, we set the flow control rule to forward all the traffic to the MIHs; on the other hand, in order to capture as many attacks as possible, we set the flow classification rule to allow any connection attempt. We used the traffic redirection mechanism 2 to deploy the virtual decoys, the well-known open-source MIH softwares, Dionaea [44] and Amun [48], were deployed in the virtual machines, and the assigned IP addresses were registered in DNS so that they can be found by attackers. Thanks to the well-developed log statistic approaches of Dionaea, the captured data can be graphically shown in Fig. 14, i.e. the accepted requests and the remote attacking hosts.

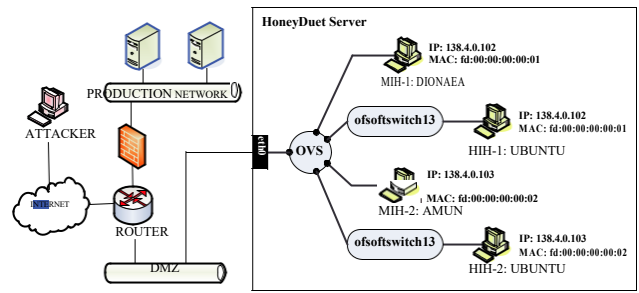

Fig. 13. Schematic diagram of system deployment in real scenario

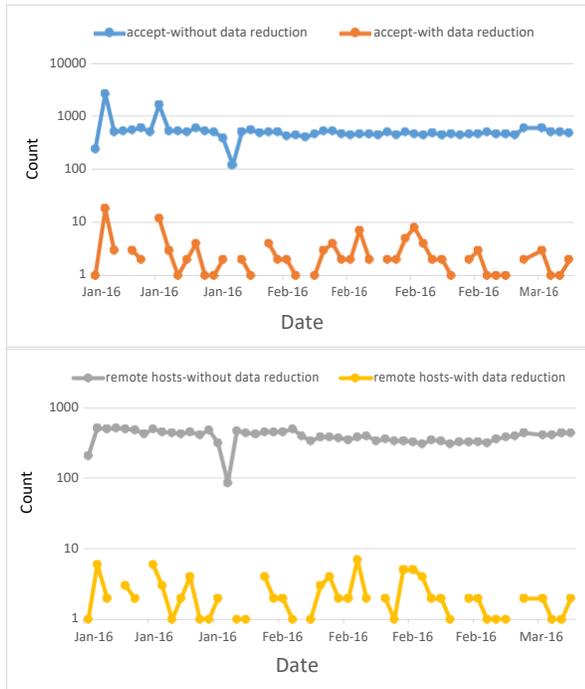

Fig. 14. Live data captured compared to the results of data reduction

During the period of live attack capture, 28099 attacking incidents were observed, and among them 169 counts hit the vulnerabilities emulated by Dionaea. The port attack frequency of these hit counts are shown in Table III.

TABLE III
Dionaea: emulated open ports and attack hit counts

| Port | 21 | 42 | 135 | 445 | 1433 | 5060 | 40950 | 42737 | 53360 |
|---|---|---|---|---|---|---|---|---|---|
| Hitcounts | 115 | 28 | 12 | 2 | 2 | 7 | 1 | 1 | 1 |

1) Sensibility utility - useless network data reduction: In order to catch data as much as possible, we did not set data filtering rules. However, we can observe the statistics presented above (Table III) that show only a small part of the data can be identified by Dionaea. Hence, if the researchers desire to get a more fine-grained traffic, a set of traffic filtering rules can be configured in order to undertake the data reduction. We set the Snort rules blocking all ports except the ones shown in Table III. Later, in order to perform the comparative experiment, we replayed the attacks in a testbed scenario with the same configuration, but the OVS is configured by the new Snort rules. After completing the attack replay, the new attack statistics of Dionaea are presented by the data reduction lines (the "accept-with data reduction" and the "remote hosts-with data reduction" lines) of Figure 14. It shows that the attack frequency decreases from several hundred times per day to several times per day. Therefore, the proposed customizable traffic classification can enforce effective data reduction.

2) Stealth utility - interesting system data capture: The data captured by the MIH is limited to conduct attack profiling and analysis, since there is no system activity left by the adversary to be used for investigation. For example, we captured a brute force attacker (with IP address 62.210.207.107) by Amun, and the recorded data can be shown as follows:

```
2015-12-26 04:38:08,031 INFO [vuln_ftpd] Attacker: 62.210.207.107 Message: ['USER anonymous \r \n'] Bytes: 16 Stage: FTPD_STAGE1
2015-12-26 04:38:08,066 INFO [vuln_ftpd] Attacker: 62.210.207.107 Message: ['PASS anonymous@ \r \n'] Bytes: 17 Stage: FTPD_STAGE1
2015-12-26 04:38:08,101 INFO [vuln_ftpd] Attacker: 62.210.207.107 Message: ['CWD / \r \n'] Bytes: 7 Stage: FTPD_STAGE2
2015-12-26 04:38:08,135 INFO [vuln_ftpd] Attacker: 62.210.207.107 Message: ['TYPE A \r \n'] Bytes: 8 Stage: FTPD_STAGE2
```

Due to the limited interaction between the attacker and the decoy (created by Amun), we observed that the captured data is readable but it provided little information. It is not like a HIH that can be compromised and allow the attacker to have a full interaction so that more attacking information can be obtained.

Hence, if we redirect the interesting traffic into the HIH, we are able to get more data about the adversary's activity for investigating the intrusion event. We emulated a real attack by using msfconsole[6] to exploit the distcc[7] vulnerability on the honeypot. In order to launch attack again the MIH, we set the redirection rule as follows:

```
alert tcp any any → any 3632 (msg:"MIH";)
```

The manually attack launched by msfconsole failed to compromise the MIH and only a TCP connection request on port 3632 was recorded. Thereby, comparing to the first redirection rule, we set to redirect the traffic to the HIH:

```
alert tcp any any → any 3632 (msg:"HIH";)
```

Thereafter, we undertook the identical attack against the distcc vulnerability on the Metasploitable2[8] honeypot again. The msfconsole successfully gained the online access. Later, we downloaded a privileged escalation exploit distcc to escalate the privilege from user daemon to root, and used the netcat send back the root shell to the attacker's side. The emulated intrusion was completed by the adversary gaining the root privilege of the compromised honeypot. Consequently, the security researcher can use the Volatility[9] scan to enforce the memory forensics over the compromised system for the purpose of attack profiling. Some forensic results can be seen in Table IV.

[6] https://www.metasploit.com/
[7] https://www.rapid7.com/db/modules/exploit/unix/misc/distcc_exec
[8] https://sourceforge.net/projects/metasploitable/files/Metasploitable2/

TABLE IV
The Volatility forensic results

| Action | Results |
| --- | --- |
| linux_netstat | New distcc Process ID |
| | New netcat Process ID |
| linux_psaux | New socket created by the metasploit payload |
| linux_pstree | Standard output file during the distcc exploit |
| | New PID that executed the shell by netcat |

## VII. Conclusion

A honeypot system is a vitally important security facility created to be probed, attacked and compromised, in order to trap the adversaries as well as investigate the well-known, and especially, the unknown attacks. The innovation of this paper is the compact honeypot architecture - HoneyDOC, which differs from the traditional honeypot architectures by using the novel Decoy-Orchestrator-Captor perspective to dissect and decouple the honeypot, enabling all-round honeypot design, which has been demonstrated by the powerful SDN-enabled architecture. By taking advantage of the SDN technology, the heterogeneous decoys supported by the SDN switches can be integrated into the versatile honeypot system flexibly, the diverse security applications can be developed and integrated upon the SDN controller's APIs, particularly, the traffic control can be adaptively and transparently conducted by the SDN controller applications according to the requirements.

A proof-of-concept system has been implemented for validating the proposal. The sensibility test shows the arbitrary traffic classification rules and the fine-grained actions. The countermeasure and stealth tests demonstrate that a much stealthier traffic migration function is added but the performance does not decrease compared to existing solutions (i.e. Honeybrid). Also, we conducted the system deploying virtual honeypots in real production network for capturing live attacks. The real data based validation shows the efficiency of data reduction and the effectiveness of the traffic redirection for data analysis. The experimental results show the feasibility and efficiency of the proposed architecture.

In the future, we will improve the whole system and use it to conduct research by long-term live data capture. We consider proposing a network functions virtualization (NFV) [51] based general honeypot deployer, to create and manage the decoys in cloud, so as to capture and analyze cyber threat data from different sources. We also take steps to enhance the anomaly-based detection, facilitate the adaptive decoy deployment to deter the advanced persistent threat (APT), and perhaps, cooperate with AI techniques to help forensics and intrusion prediction.

## Acknowledgment

The skills, knowledge and experience W.Fan gained from the C3ISP project (the European Union's Horizon 2020 research and innovation programme under grant agreement No 700294) he participated in are very helpful to complete the work.

---

[9] http://www.volatilityfoundation.org/